\documentclass[twocolumn,pre,A4paper]{revtex4}

\usepackage{epsfig,graphicx}
\newcommand{\eps}{\varepsilon}

\begin{document}

\title{Lamellar Phases in Nonuniform Electric Fields: Breaking the
In-Plane Rotation Symmetry and the Role of
Dielectric Constant Mismatch}

\author{Y. Tsori\\
Department of Chemical Engineering, \\
Ben Gurion University of the Negev,\\
Beer Sheva 84105, Israel.\\
Email: tsori@bgu.ac.il}

\vspace{1cm}

\date{28/1/2007}

\begin{abstract}
We consider orientational transitions of lamellar phases under the influence of
a spatially nonuniform
electric field. The transition between parallel and perpendicular lamellar
stackings with respect to the substrate is investigated as a function of the
system parameters. 
The dielectrophoretic energy and the energy penalty for having
dielectric interfaces perpendicular to the field's direction are identified as
linear and quadratic terms in a free energy expansion in the dielectric
constant mismatch.
We find that if the dielectric constant mismatch $\Delta\eps$ is smaller than some critical
value $\Delta\eps_c$, parallel lamellar stacking will be realized, no matter how large the
voltage difference between electrodes is.
At $\Delta\eps>\Delta\eps_c$, perpendicular stacking will appear if the voltage is high enough.
Nonuniform fields remove the in-plane degeneracy present in
the more
common uniform fields. We therefore calculate the energy of grains of different
orientations. The torque acting on the grains leads to the preference of only
one orientation. 
The results have direct implications to block copolymer
orientation and to surface patterning on the
nanometer scale.
\end{abstract}

\maketitle

\noindent {\bf Introduction}. In recent years we have seen a large effort directed toward finding ways to
control the phase-behavior and orientation of self-assembled structures
\cite{thomas_rev,av_leibler}. Confinement between two solid surfaces
\cite{russell1,
nealey,turner92,turner94,turner97,muthu1,TA1,TJ,hash2,TA3,tash} , shear flow
\cite{GHF}, or the use of external electric fields
\cite{AH93,russell2,TA2,muthu2,russell3,schick,krausch1,krausch2,russell4,
russell5,matsen,zvelin1,zvelin2} have
proved very useful. The
use of electric fields is especially appealing, as the field strength scales
favorably with the system size. Spatially uniform electric fields, however, pose
a long-lasting problem since the orientation of the assembled phases is not
unique -- the
symmetry of the field means that all grain rotations in the plane are
energetically equivalent. 

\begin{figure}[h!]
\begin{center}
\includegraphics[scale=0.4,bb=15 290 570 800,clip]{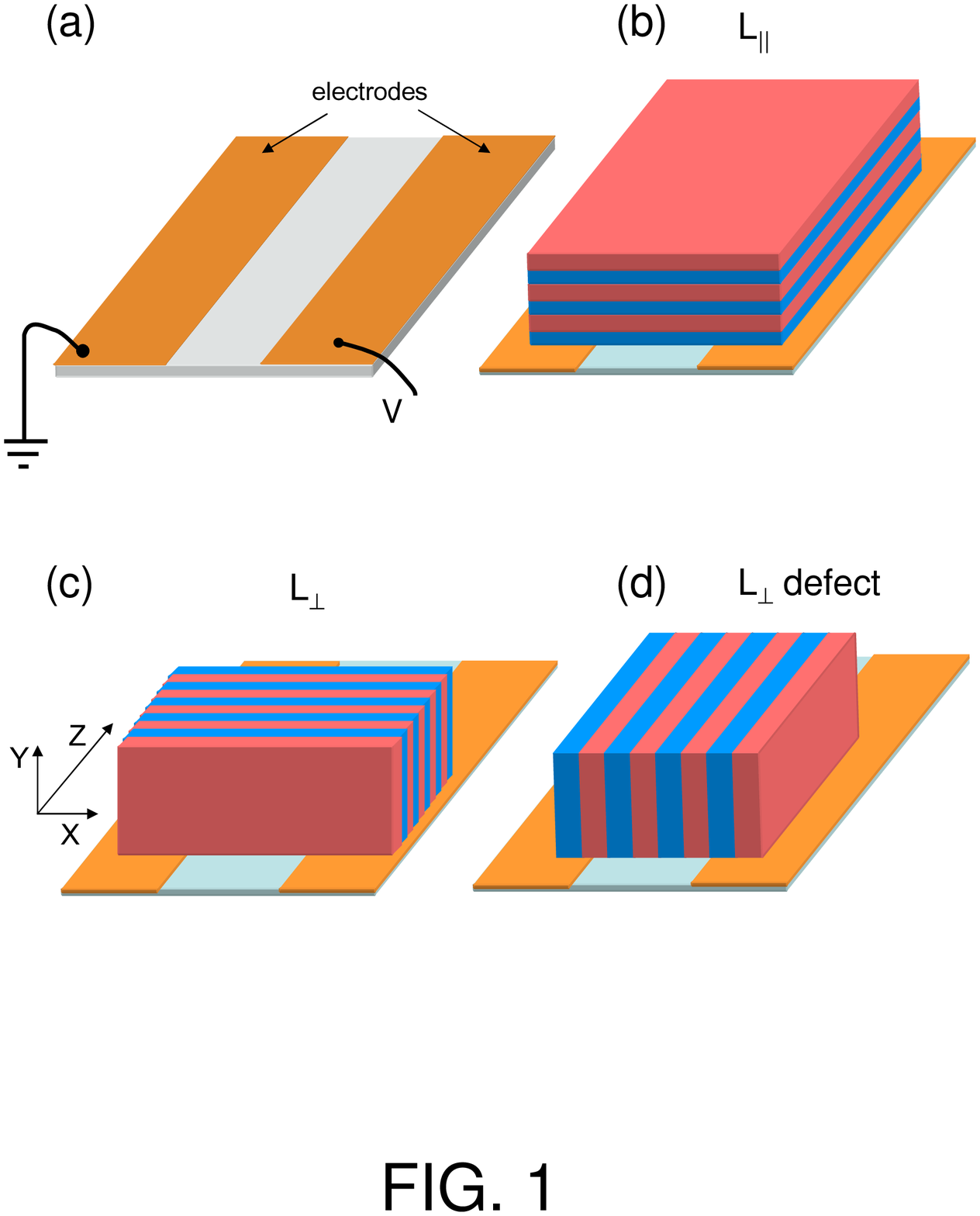}
\end{center}
\caption{\footnotesize{Schematic illustration of the system. (a) Two thin
 ``razor-blade'' electrodes are laid down on the substrate. The voltage
 difference between them is $V$. (b) Parallel stacking $L_\parallel$ -
 lamellae lie parallel to the substrate. (c) If the voltage is
 sufficiently high, electric field can overcome interfacial interactions
 and prefer a perpendicular stacking $L_\perp$ (lamellae are parallel to
 the field lines). (d) A defect -- an unfavorable perpendicular morphology
 where lamellae are perpendicular to the field lines. In subsequent
 calculations we took the distance between electrodes to be $1~\mu$m, and
 the lamellar period is $100$ nm unless otherwise indicated.
}}
\end{figure}

Spatially varying fields remove this degeneracy, and thus can be quite useful in
alignment of various mesophases. The early experiments of Russell {\it et al.} \cite{russell7} have employed nonuniform fields, but ever since then all research have been on uniform fields. It seems that now, when such spatially 
uniform fields have been well understood and exploited possibly to their full 
potential, it is time to come back to spatially varying fields.
In this article we focus on the most simple
periodic structure -- the lamellar phase, which is found under the influence of
an electric field emanating from a ``razor-blade'' electrode design (see Fig. 1).
The lamellae are made up of two different materials, A and B, e.g., diblock
copolymers. In this example, the two polymers A and B have different dielectric
constants, $\eps_B$ and $\eps_B$. In the following we assume ion-free polymers;
alternatively, for ion-containing polymers, application of a quasi-static field
in the frequency $\sim1$ kHz renders the ions immobile but leaves the
electrostatic equations unchanged \cite{TA2}. 
In spatially uniform electric fields, the lowest-order contribution to the
system electrostatic free energy is quadratic in the permittivity difference
of the two constituents, $\Delta\eps\equiv \eps_A-\eps_B$.
As is explained in detail below, an inhomogeneity of the field gives rise to
a dielectrophoretic force which is manifested in a linear term in $\Delta\eps$, and this has significance to the orientation
selection and to phase-transitions \cite{TTL}.

At this point it should also be pointed out that nonuniform electric fields are in
general neither
interfacial nor purely bulk ones. In the razor-blade geometry, the field is 
high close to the electrodes' edge. However, sufficiently far from the electrodes the field
behaves like $E(r)=V/\pi r$, where $r$ is the distance from the middle of
the gap. Thus, the integrated 
electrostatic contribution to the energy scales like $~1/r$. This energy indeed
decays, but very slowly, and it has an important contribution even very
far from the electrodes.

We assume that the lamellae are rigid enough so that the electric field does
not
bend them. In the example of block copolymers, this corresponds to the so-called
strong-segregation regime, where $\chi N\gg 1$. Let us verify the validity of this
assumption. The elastic bending energy per unit volume is written as
$F_{\rm el}=\frac12K/R^2$, 
Where $K$ is the bending modulus, and $R$ is the local bending radius
(inverse curvature). For strongly stretched lamellae, $K=D\gamma_{AB}$, where $D$
is the lamellar period ($D\sim 100$ nm) and $\gamma_{AB}\sim 10$ mN/m is the A/B
interfacial tension. 
On the other hand, the electrostatic energy per unit volume is $F_{\rm
es}=\frac12\eps E^2$,
where $\eps$ is the dielectric constant and $E$ the local field, which cannot 
exceed $\sim 100$ V/$\mu$m because of dielectric breakdown. Let us take this maximum
value, in this case $F_{\rm es}=10^5$ J/m$^3$.
Therefore, $F_{\rm el}=F_{\rm es}$ if the lamellae are bent with a radius of curvature of
$0.1~\mu$m. The same estimate relates to the
stresses (forces) of course. Electric fields cannot bend lamellae to a radius smaller than
$\sim 0.1~\mu$m.
In the razor-blade system, at larger distances from the electrodes the field is weaker,
and therefore the lamellae should stay flat as well. Since the fields we consider are
typically much weaker, we do not expect 
bent lamellae in this electrode
arrangement. The above reasoning does not hold for weakly segregated lamellae:
These lamellae have a much weaker modulus $K$ and therefore significant bending can
occur. 

A lamellar stack can therefore
have the basic configurations: parallel or perpendicular to the substrate
(Fig. 1 (b) and (c)), denoted as $L_\parallel$ and $L_\perp$, respectively.
Note that in the parallel stacking the first layer at the substrate is half as
thick as the others. A
third state exists which we denote the perpendicular defect. Here the lamellae
normals are not parallel to the electrodes' edges. Fig. 1 (d) represents the
highest energy of such defects. 
Weakly segregated systems exhibit lower energy defects, e.g., 
T-junction or grain boundary. In some experiments with weakly segregated
block copolymers on preferential surfaces, few lamellar layers are adsorbed
preferentially on the substrate (mixed morphology) due to the
long-range effect of surface ordering \cite{TA2,russell6}. 
As is mentioned above, this system is out of the scope of the
current work, and it will be dealt with in a subsequent publication.

A peculiar feature of nonuniform fields is that the $L_\parallel$ state can be
favored over the $L_\perp$ one even in the absence of specific interfacial
interactions with the substrate. In order to understand this, consider first
the distribution of electric field squared for two semiinfinite planar
electrodes in the
$x$--$z$ plane, with a gap of $1$ $\mu$m between them. This distribution is
shown in Fig. 2 for a medium with spatially uniform dielectric constant $\eps$.
Clearly $E^2$ is very high close to the surface and, in particular, close to the
electrodes' edge at $x=\pm 0.5$ $\mu$m. The field is small far from the
substrate, and therefore interfacial instabilities are not expected \cite{matsen_prl}; 
this is
true even more so since above the electrodes' edges the field at $y\to\infty$ is
actually parallel to the substrate and also to the polymer/air interface.

Let us now assume
without loss of generality that $\eps_A>\eps_B$. As is well-known
in the field of dielectrophoretic forces \cite{pohl}, a material with large
value of $\eps$
is drawn to regions with high fields, whereas small-$\eps$ material is
repelled.
Since the electric field is
largest near the electrodes' edges, an $L_\parallel$ state can form with the
A-material touching the substrate. However, the work of Amundson
{\it et al.} has shown that there is also a free energy
penalty for having dielectric interfaces perpendicular to the field's
direction, and this penalty is absent in the $L_\perp$ state. Clearly, the
orientation selection depends on the magnitude of $\eps_A-\eps_B$.
\begin{figure}[h!]
\begin{center}
\includegraphics[scale=0.45,bb=40 240 550 530,clip]{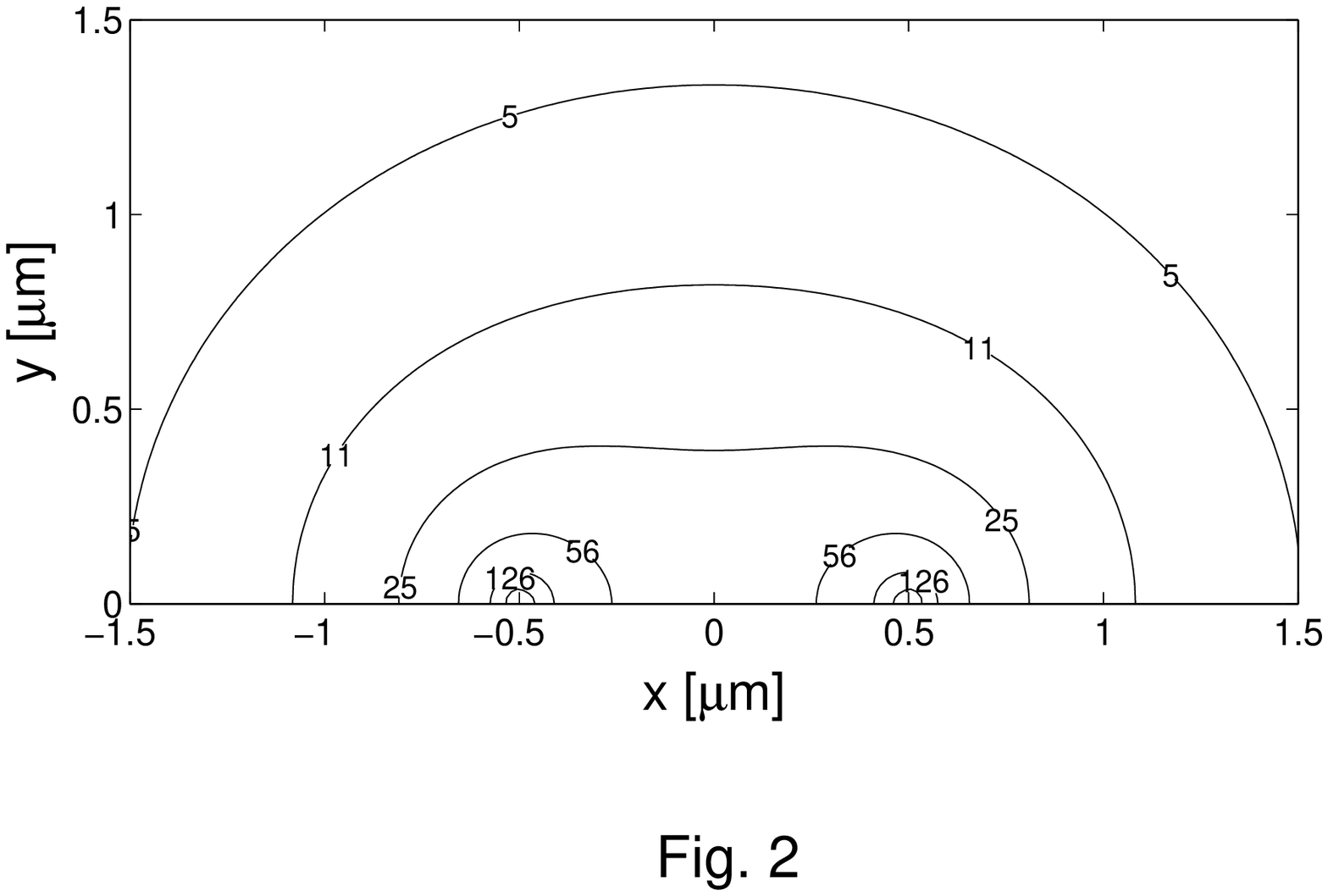}
\end{center}
\caption{\footnotesize{Plot of $E^2(x,y)$ in the $x$--$y$ plane, for the
case where the dielectric constant $\eps$ is uniform, and the electrodes are at
$x>0.5$ $\mu$m ($V=\frac12$ V) and $x<-0.5$ $\mu$m ($V=-\frac12$ V). 
The largest field is
at the electrodes' edge, $x=\pm 0.5$ $\mu$m.
$E^2$ is scaled by $10^{10}$ and given in (V/m)$^2$.
}}
\end{figure}

The electrostatic energy of the system is given by an integral over all space,
\begin{eqnarray}\label{Fes}
F_{\rm es}=-\frac12\int\eps({\bf r}){\bf E}^2({\bf r})~{\rm d}^3r 
\end{eqnarray} 
The dielectric constant $\eps({\bf r})$ is a spatially varying quantity. In
this study it is a periodic function. In the $L_\parallel$ state, for example,
it is given by
\begin{eqnarray}\label{sq_wave}
\eps({\bf r})=\left\{
\begin{array}{c}
\bar{\eps}+\frac12\Delta\eps ~~~~~{\rm
if}~~nd<y<nd+\frac12 d,~~~~~~~~~n=0,1, 2 ...\\
~\\
\bar{\eps}-\frac12\Delta\eps ~~~~~{\rm
if}~~nd+\frac12<y<(n+1)d,~~~n=0,1, 2 ...
\end{array}\right.
\end{eqnarray} 
where $\bar{\eps}\equiv \frac12(\eps_A+\eps_B)$ is the average dielectric
constant, and the period is $d$.
The above equation simply represents a square-wave in the $y$-direction, where
$\eps$ alternates between $\eps_A$ and $\eps_B$. The dielectric constant
can be defined similarly for the other stackings. 

~\\
\noindent {\bf Theory and Results}.
Figure 3 shows $F_{\rm es}$ for the $L_\parallel$ and $L_\perp$ stackings at a
fixed value of $\bar{\eps}=6$ and varying values of the dielectric constant
mismatch. The electrostatic energy is calculated numerically for a system with
electrode gap of $1~\mu$m. $F_{\rm es}(\Delta\eps)$ (dashed horizontal line) is
constant for the
$L_\perp$ case, because the electric field between the electrodes is independent
of
$\Delta\eps$ and $\bar{\eps}$. On the other hand, in the $L_\parallel$
case (solid line), $F_{\rm es}(\Delta\eps)$
decreases first before it increases. The decrease is due to the
dielectrophoretic term, linear in $\Delta\eps$, while the increase is due to
the penalty associated with dielectric interfaces perpendicular to the field
lines, scaling like $(\Delta\eps)^2$.

Let us make a short but very general mathematical digression which will clarify
the last point. Denote ${\bf E}_0({\bf r})$ the electric field which corresponds
to a system of uniform dielectric constant and a given electrode
design (not necessarily the one in Fig. 1). ${\bf E}_0$ is derived from a
potential $\psi_0({\bf r})$ satisfying the proper boundary conditions on the
electrodes: ${\bf E}_0=-\nabla\psi_0$. Suppose now that the dielectric constant
changes from its average value by an amount $\eps_1({\bf r})$: $\eps({\bf
r})=\bar{\eps}+\eps_1({\bf r})$. This change in permittivity leads to a change
in field: ${\bf E}({\bf r})={\bf E}_0({\bf r})+{\bf E}_1({\bf r})$. We may now
write the integrand of Eq. (\ref{Fes}) in the following way:
\begin{eqnarray}\label{Fes_exp}
f_{\rm es}&\equiv &-\frac12\eps{\bf E}^2=-\frac12\bar{\eps}{\bf
E}_0^2-\frac12\left[\eps_1{\bf E}_0^2+2\bar{\eps}{\bf E}_0\cdot{\bf
E}_1\right]\nonumber\\
&-&\frac12\left[2\eps_1{\bf E}_0\cdot{\bf E}_1+\bar{\eps}{\bf
E}_1^2\right]-\frac12\eps_1{\bf E}_1^2
\end{eqnarray} 
The first term on the right is the electrostatic energy of the system with
uniform average $\eps$, while the other three terms are the deviations from it.
The second and third terms (square brackets) are the dielectrophoretic and
``dielectric interfaces'' terms, scaling like $\eps_1$ and $\eps_1^2$,
respectively. Finally, the last term scales like $\eps_1^3$, and is small if
$\eps_1\ll\bar{\eps}$. For the case where this last term is dealt with the
interested reader is referred to \cite{schick}.
\begin{figure}[h!]
\begin{center}
\includegraphics[scale=0.5,bb=50 260 485 580,clip]{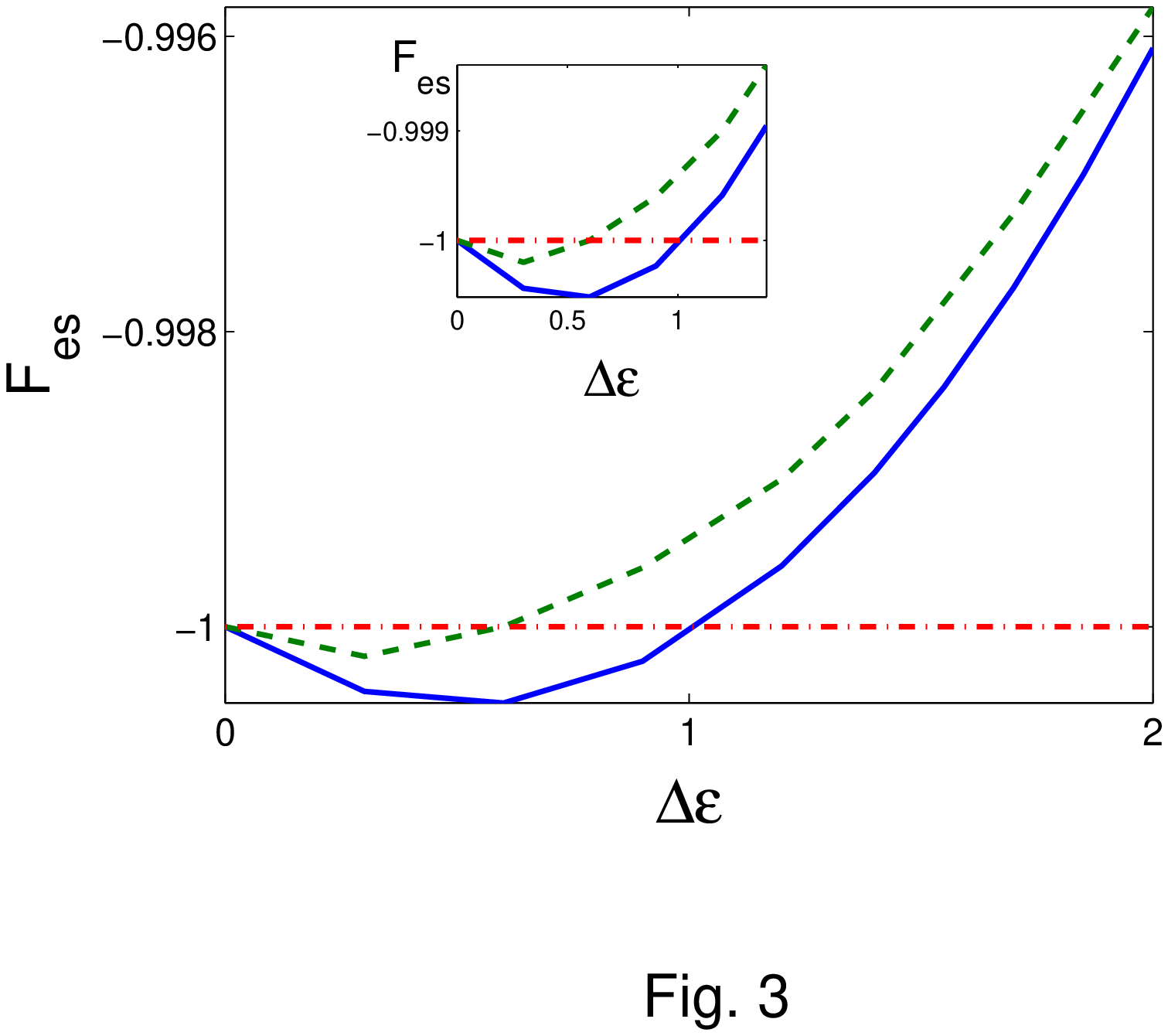}
\end{center}
\caption{\footnotesize{Numerically calculated electrostatic energy
$F_{\rm
es}$ [Eq. (\ref{Fes})] of parallel $L_\parallel$ (solid line) and
perpendicular $L_\perp$ (horizontal dash-dotted line) stackings as a
function of the permittivity difference:
$\Delta\eps\equiv\eps_A-\eps_B$. $F_{\rm es}$ is normalized
by its value when $\Delta\eps=0$. $F_{\rm es}$ of perpendicular lamellae is
constant, while that of parallel ones decreases before it increases (see inset).
The critical value of
$\Delta\eps$ is $\Delta\eps_c\simeq1$. When $\Delta\eps<\Delta\eps_c$,
$L_\parallel$ is preferred over $L_\perp$. If
$\Delta\eps>\Delta\eps_c$, $L_\perp$ is preferred.  
We took the average dielectric constant to be $\bar{\eps}=6$, the lamellar period is $100$ nm, and the electrode gap is $1$ $\mu$m. The dashed line is a
similar plot of
$F_{\rm es}$ for $L_\parallel$ lamellae, with the same parameters; only the electrode gap is $2$ $\mu$m.
}}
\end{figure}

On the basis of this expansion and denoting ${\bf E}_1=-\nabla\psi_1$, one can easily
show that $\psi_1$ obeys the following
equation
\begin{eqnarray}
\nabla^2\psi_1=\frac{1}{\bar{\eps}}\nabla\eps_1\cdot{\bf E}_0
\end{eqnarray} 
with the boundary conditions that $\psi_1=0$ on all conductors. 
Clearly $\psi_1$ can be written as $\psi_1=\psi_1({\bf r},\eps_1/\bar{\eps},{\rm
geometry},V)$, where geometry refers to the electrode geometry and $V$ to the
electrode potential difference (in the case of just two electrodes).
We now write
$\eps_1$ in a form that puts emphasis on dimensions: $\eps_1({\bf
r})=\Delta\eps\cdot c({\bf r})$. Thus, $c({\bf r})$ is a dimensionless function
containing the spatial variation of $\eps_1$, and whose spatial average
vanishes:
$\langle c({\bf r})\rangle=0$. For the square-wave example of
Eq. (\ref{sq_wave}), $c=\pm\frac12$.
It then directly follows that 
\begin{eqnarray}
\psi_1=\frac{\Delta\eps}{\bar{\eps}}\tilde{\psi}_1({\bf r};c({\bf r}),{\rm
geometry},V)
\end{eqnarray} 
\begin{figure}[h!]
\begin{center}
\includegraphics[scale=0.45,bb=5 260 490 580,clip]{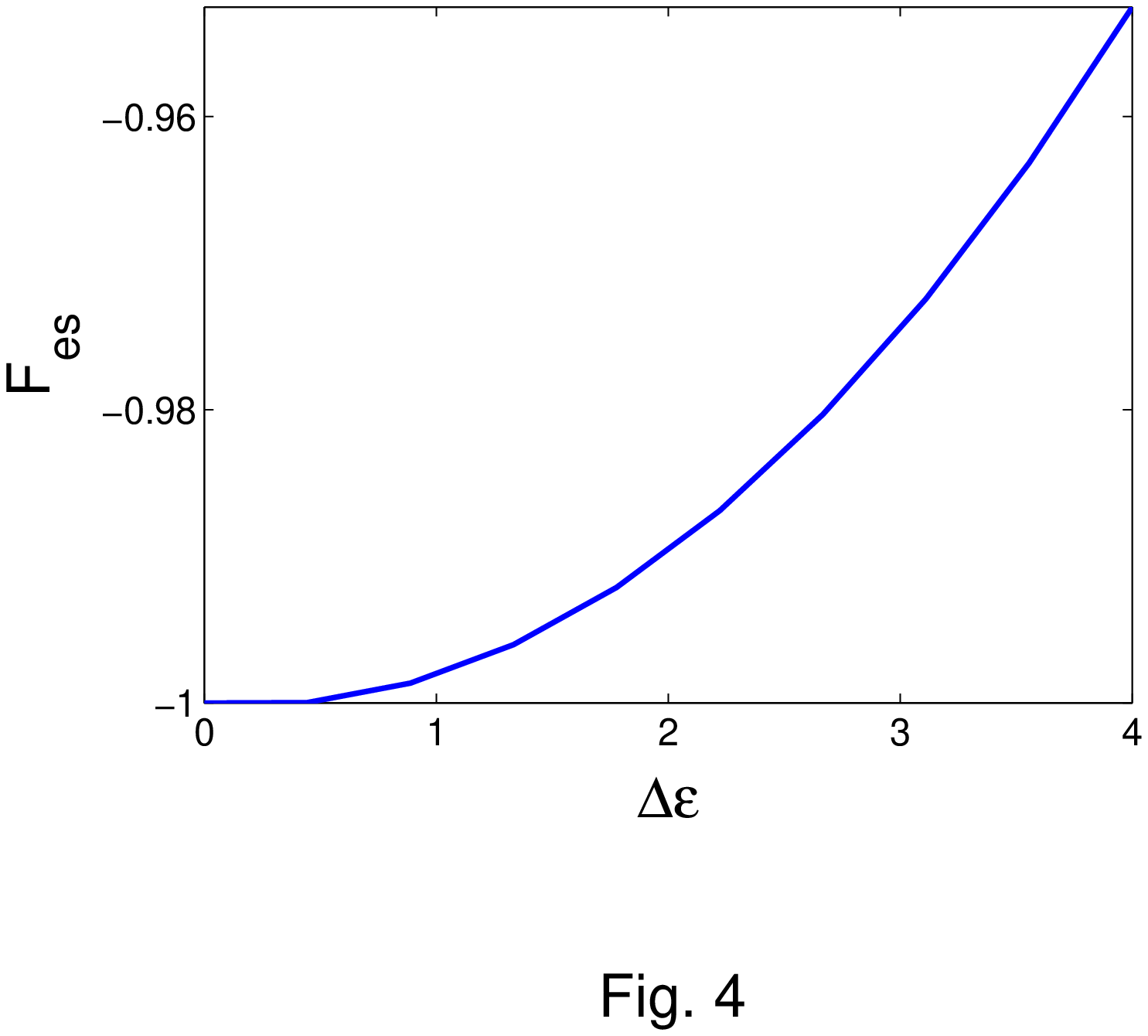}
\end{center}
\caption{\footnotesize{Electrostatic energy $F_{\rm es}$ of perpendicular-defect
structure (Fig. 1 (d)) as a function of $\Delta\eps$. $F_{\rm
es}$ is normalized by its value
when $\Delta\eps=0$ and is always increasing. Other parameters as in Fig. 3
}}
\end{figure}
where $\tilde{\psi}_1$ obeys the equations
\begin{eqnarray}
\nabla^2\tilde{\psi}_1=\nabla c\cdot{\bf E}_0
\end{eqnarray} 
and $\tilde{\psi}=0$ on all electrodes. Since $\tilde{\psi}_1$ is a universal
potential independent of $\Delta\eps$, $\psi_1$ is linear in
$\eps_1/\bar{\eps}$ (and in fact it is linear in $V$ as well). Similarly we find
${\bf E}_1=(\Delta\eps/\bar{\eps})\tilde{{\bf E}}_1({\bf r};c({\bf
r}),{\rm
geometry},V)$, with $\tilde{\bf E}_1$ independent of $\Delta\eps$.
We now rewrite Eq.
(\ref{Fes_exp}) as follows:

\begin{eqnarray}\label{lin_quad_deps}
F_{\rm es}&=&\Delta\eps I_1+\frac{(\Delta\eps)^2}{\bar{\eps}}I_2+const.\\
I_1&=&-\frac12\int \left[c({\bf r}){\bf
E}_0^2+2{\bf E}_0\cdot{\bf \tilde{E}}_1\right]~{\rm d}^3r\nonumber\\
I_2&=&-\frac12\int\left[2c({\bf r}){ \bf E}_0\cdot{\bf \tilde{E}}_1+{\bf
\tilde{E}}_1^2\right]
~{\rm d}^3r\nonumber
\end{eqnarray} 
The expansion of $F_{\rm es}$ is now transparent to order $(\Delta\eps)^2$, as
both $I_1$ and $I_2$ are independent of $\Delta\eps$, are quadratic in $V^2$, 
and depend on geometry and $c({\bf r})$. 
In order to further demystify the above expansion, consider the simple 
one-dimensional example of uniform electric field $E_0$ (parallel-plate
capacitor), with $c=\pm\frac12$. In this case we find $\tilde{E}_1=-cE_0$ and
$E_1=-c(\Delta\eps/\bar{\eps})E_0$, and since $\langle c\rangle=0$ we find
a rather well-known result: $\langle f_{\rm
es}\rangle=\frac18[(\Delta\eps)^2/\bar{\eps}]
E_0^2+const$.
\begin{figure}[h!]
\begin{center}
\includegraphics[scale=0.35,bb=90 170 530 800,clip]{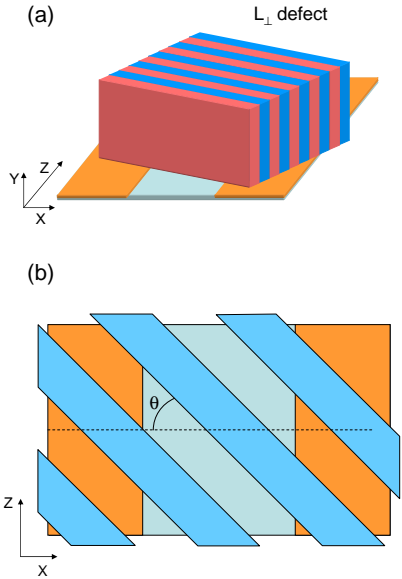}
\end{center}
\caption{\footnotesize{(a) Illustration of a defect perpendicular morphology.
Lamellae
make an angle
$\theta$ in the $x$--$z$ plane, as defined in (b). The system experiences
torque which tends to align the stacking, preferring the state with
$\theta=0$.
}}
\end{figure}

We now return to the razor-blade electrode design and the results presented in
Fig. 3. The descent of $F_{\rm es}$ for parallel lamellae 
is due to a negative value of $I_1$,
stemming from the dielectrophoretic force. The subsequent increase at larger
value of $\Delta\eps$ is due to a positive $I_2$. The critical value of
$\Delta\eps$, $\Delta\eps_c$, is given by the relation 
\begin{eqnarray}
\Delta\eps_c=-\bar{\eps}I_1/I_2
\end{eqnarray} 
The existence of $\Delta\eps_c$ is indeed important -- at all $\Delta\eps<\Delta\eps_c$ the
morphology is that of parallel layers ($L_\parallel$), {\it irrespective of the
applied voltage} or the magnitude of the electric field. In uniform electric
fields similar critical value of $\Delta\eps$ does not exist. The value of the
last term ignored in Eq. (\ref{Fes_exp}) is numerically verified to be
negligible in this calculation.
\begin{figure}[h!]
\begin{center}
\includegraphics[scale=0.45,bb=50 250 490 575,clip]{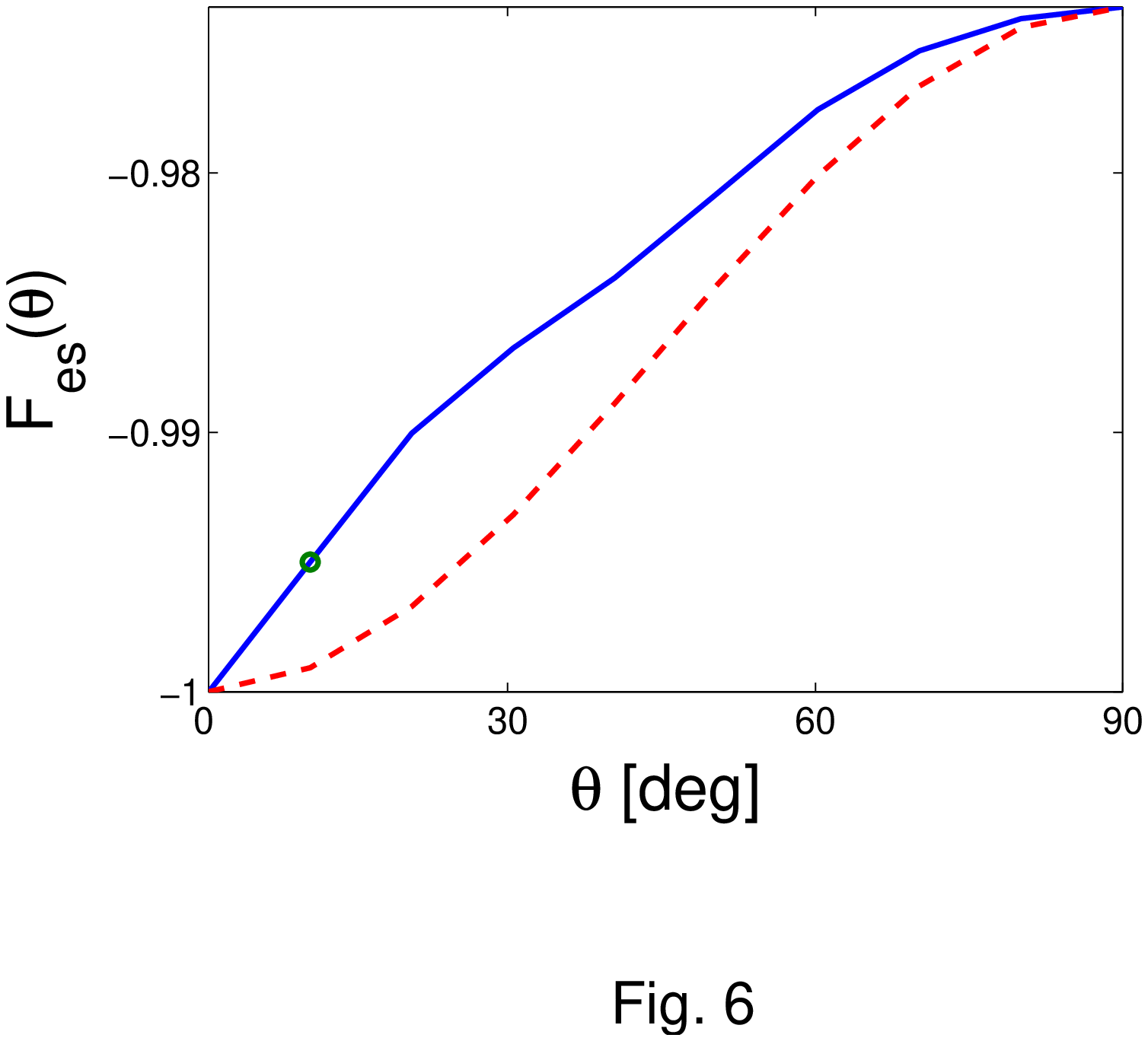}
\end{center}
\caption{\footnotesize{Solid line: electrostatic energy $F_{\rm es}$ of
perpendicular
lamellae as a
function of rotation angle $\theta$ defined in Fig. 5. $F_{\rm es}$ is
scaled by $|F_{\rm es}(\theta=0)|$.  $\theta=0$ corresponds to ``perfect''
perpendicular
layering, while $\theta=90^\circ$ is the defect with the highest energy.
The torque is $L=dF_{\rm es}/d\theta$. Dashed line: a fit interpolating
the maximum and minimum values by a $\sin^2(\theta)$ fit: $F_{\rm
es}=F_{\rm es}(0)+\left[F_{\rm es}(90^\circ)-F_{\rm
es}(0)\right]\sin^2(\theta)$. We took $\eps_A=8$ and
$\eps_B=4$, yielding $\Delta\eps=4$ and $\bar{\eps}=6$. The numerical accuracy
for the point marked with a circle is questionable.
}}
\end{figure}

In Fig. 4 we plot $F_{\rm es}$ as a function of $\Delta\eps$ for the
perpendicular-defect state sketched in Fig. 1 (d). At a given voltage and
$\Delta\eps$, this state has the highest electrostatic energy since the two
electrostatic terms are unfavorable -- the electrodes are not
covered with the
high-$\eps$ material ($I_1>0$), and the field lines cross the lamellar
interfaces ($I_2>0$).

Figure 5 depicts a lamellar grain in a defect state: the lamellae normals are
not parallel to the electrodes' edges. The highest energy rotation has
$\theta=90^\circ$, while the lowest is the $L_\perp$ state with $\theta=0$.
In Fig. 6 we present the electrostatic energy $F_{\rm es}$ as a function of the
rotation angle $\theta$. The torque acting on the sample to orient it in the
preferred direction is given as the derivative: $L=dF_{\rm
es}(\theta)/d\theta$; it vanishes for the two extreme cases
$\theta=0$ and $\theta=90^\circ$ \cite{TA3,TA2}. Indeed, when ${\bf E}_0$ is
uniform in space we find 
$F_{\rm es}(\theta)=F_{\rm es}(0)+\left[F_{\rm es}(90^\circ)-F_{\rm
es}(0)\right]\sin^2(\theta)$. As is seen in the figure, the actual energy is
higher than this estimate.
\begin{figure}[h!]
\begin{center}
\includegraphics[scale=0.45,bb=70 260 490 580,clip]{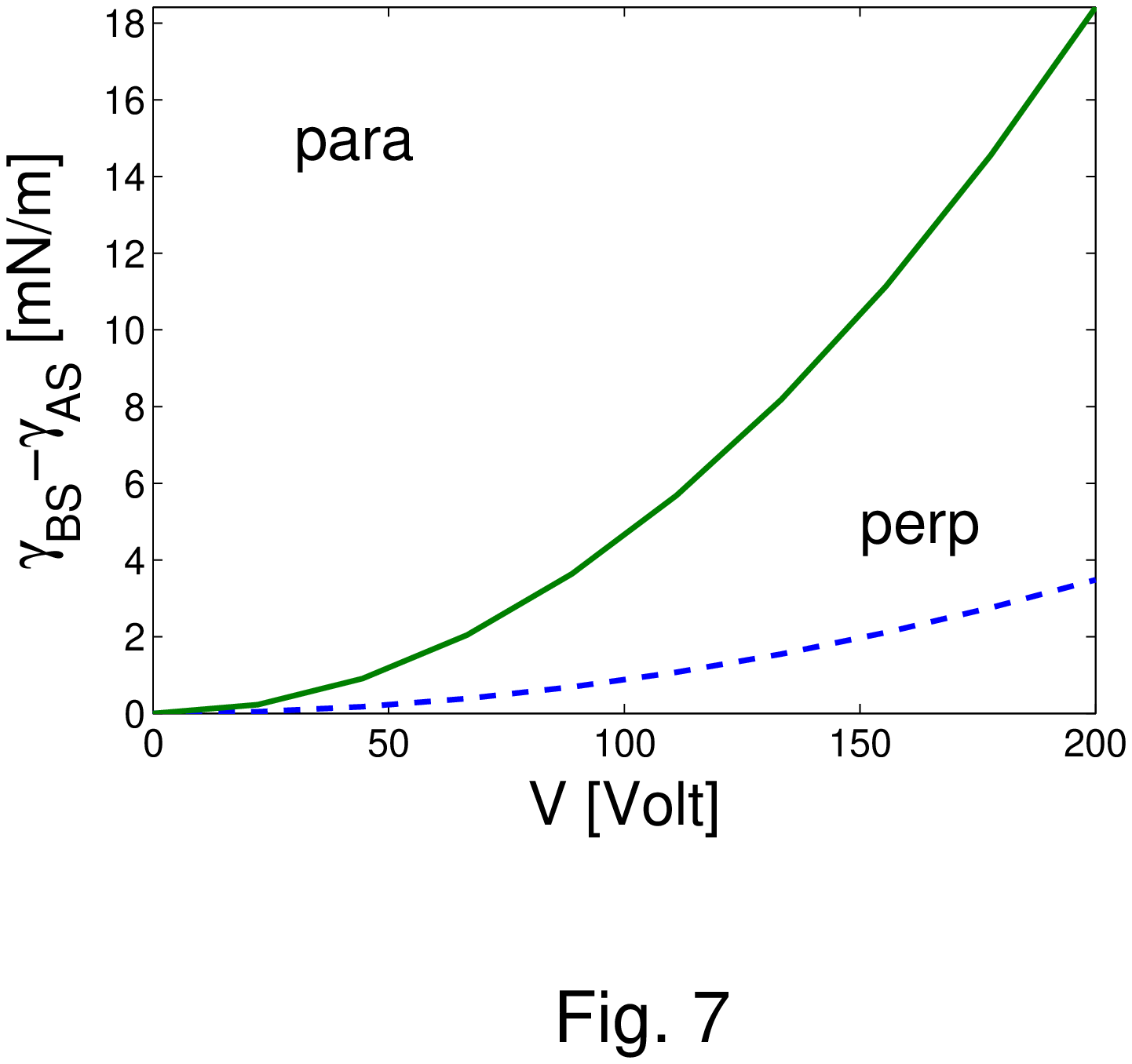}
\end{center}
\caption{\footnotesize{Phase diagram in the voltage--interfacial interactions
plane. $V$ is the
voltage between the electrodes (see Fig. 1), and $\gamma_{AS}$ and
$\gamma_{BS}$ are the interfacial interactions of the A and B polymers
with the substrate. Above the solid line (green) and for $\Delta\eps=4$,
$L_\parallel$ is stable, while below it $L_\perp$ is expected. The dashed
blue line is the same, but for $\Delta\eps=2$. In both cases
$\bar{\eps}=4$, $\Delta\eps>\Delta\eps_c\simeq 1$, $d=100$ nm, and the electrode
gap is $1 \mu$m.
}}
\end{figure}

Finally, the interfacial interaction of the two materials with the substrate
can be taken into account as well. Let us call $\gamma_{\rm AS}$ and
$\gamma_{\rm BS}>\gamma_{\rm AS}$ the interfacial energies per unit area of the
A and B materials with the surface, respectively. The free energy difference 
between the $L_\parallel$ and $L_\perp$ states is
\begin{eqnarray}
\Delta F=I_1\Delta\eps+I_2\frac{(\Delta\eps)^2}{\bar{\eps}}+\frac12
S(\gamma_{\rm AS}-\gamma_{\rm BS})
\end{eqnarray} 
where $S$ is the substrate area. The prevailing state is $L_\parallel$ if
$\Delta F$ is negative and $L_\perp$ otherwise. On the basis of this free energy
difference, one can construct a phase diagram, which is shown in Fig. 7 for two
values of $\Delta\eps$. Note that both $I_1$ and $I_2$ are proportional to 
$V^2$,
and since $\Delta\eps>\Delta\eps_c$, the electric field terms favor the perpendicular
stacking.
For fixed interfacial interactions, raising the voltage from small
values to large ones destabilizes the $L_\parallel$ and leads to perpendicular
stacking $L_\perp$. The critical voltage for this transition scales like
$(\gamma_{\rm AS}-\gamma_{\rm AS})^{1/2}$.

The polymer melt can be confined by another solid surface from the top. In this case
there are two more $\gamma_{AS}$ and $\gamma_{BS}$ corresponding to the second surface,
and the augmented version of the equation above reads
\begin{eqnarray}
\Delta F&=&I_1\Delta\eps+I_2\frac{(\Delta\eps)^2}{\bar{\eps}}+\frac12
S(\gamma_{\rm AS_1}-\gamma_{\rm BS_1})\nonumber\\
&+&\frac12
S(\gamma_{\rm AS_2}-\gamma_{\rm BS_2})
\end{eqnarray} 
where the ``1'' and ``2'' subscripts refer to the bottom and top surface respectively.
Here we have assumed that the film is sufficiently thick so that the
incommensurability between the lamellar thickness and the surface separation can
be neglected and the parallel lamellae are not frustrated, as is the case for surface
separation larger than $\sim 10$ lamellae.

~\\
\noindent {\bf Conclusions}. Lamellar phases under the influence of a spatially
nonuniform electric field are considered. The role of the dielectric constant mismatch
$\Delta\eps$ is highlighted: the linear term in the free energy expansion is
due to a dielectrophoretic force, while the quadratic term includes the
free energy penalty for having dielectric interfaces perpendicular to the
field's direction. We have shown that a simple electrode realization which
gives rise to nonuniform fields can bring about orientational transitions
between several lamellar stackings. Specifically, for $\Delta\eps<\Delta\eps_c$,
parallel lamellae are preferred over perpendicular ones even at very high
voltages. When
$\Delta\eps>\Delta\eps_c$, there is an interplay between electrostatic forces
and interfacial interactions. The ``razor-blade'' electrode design suggested
here can find numerous applications in nanotechnology: the large torque
is expected to
remove the degeneracy between the $L_\perp$ states by orienting the lamellae
perpendicular to the substrate and the electrodes' edges.
More complex morphologies are expected to occur for block copolymers in the intermediate
and weak segregations where the lamellar bending and grain boundary energies are smaller,
and these systems should be systematically explored in this and more advanced electrode arrangements.

\section*{Acknowledgments}

Numerous discussions with D. Andelman, L. Leibler, V.
Olszowka, T. P. Russell, A. V. Ruzette, M. Schick, H. Schoberth, K. Schmidt, 
and F. Tournilhac are
gratefully acknowledged. I am indebted to A. B\"{o}ker and G. Krausch for
several discussions and for communicating to me the results of unpublished work.
This research was supported by the Israel Science Foundation (ISF) under
Grant No. 284/05.


\begin{thebibliography}{99}

\bibitem{thomas_rev} Park, C.; Yoon, J.; Thomas, E. L. {\it
Polymer} {\bf 2003}, {\it 44}, 6725.

\bibitem{av_leibler} Ruzette, A. V.; Leibler, L. {\it Nat. Mater.} {\bf 2005}, {\it 4}, 19.

\bibitem{russell1} Mansky, P.; Russell, T. P.; Hawker, C. J.; Mayes, J.;
Cook, D. C.; Satija, S. K. {\it Phys. Rev. Lett.} {\bf 1997}, {\it 79}, 237.

\bibitem{nealey} Wang, Q.; Yan, Q.; Nealey, P. F.; de Pablo, J. J.
{\it J. Chem. Phys.} {\bf 2000}, {\it 112}, 450.

\bibitem{turner92} Turner, M. S. {\it Phys. Rev. Lett.} {\bf 1992}, {\it
69}, 1788.

\bibitem{turner94} Turner, M. S.; Rubinstein, M.; Marques, C. M.; {\it
Macromolecules} {\bf 1994},
{\it 27}, 4986. Turner, M. S.; Maaloum, M.; Ausserr\'e, D.;
Joanny, J.-F.; Kunz, M.
{\it J. Phys. II} {\bf 1994},
{\it 4}, 689.

\bibitem{turner97}  Li, Z.; Qu, S.; Rafailovich, M. H.; Sokolov, J.;
Tolan, M.; Turner, M. S.; Wang, J.; Schwarz, S. A.; Lorenz, H.;
Kotthaus, J. P. {\it Macromolecules} {\bf 1997}, {\it 30}, 8410.

\bibitem{muthu1} Petera, D.; Muthukumar, M. {\it J. Chem. Phys.} {\bf
1998}, {\it 109}, 5101.

\bibitem{TA1} Tsori, Y.; Andelman, D. {\it J. Chem. Phys.} {\bf 2001},
{\it 115}, 1970. Tsori, Y.; Andelman, D. {\it Eur. Phys. J. E}
{\bf 2001}, {\it 5}, 605.

\bibitem{TJ} Turner, M. S.;  Joanny, J.-F. {\it Macromolecules}
{\bf 1992}, {\it 25}, 6681.

\bibitem{hash2} Sivaniah, E.; Hayashi, Y.; Matsubara, S.; Kiyono, S.;
Hashimoto, T.; Fukunaga, K.; Kramer, E. J.; Mates, T. {\it
Macromolecules} {\bf 2005}, {\it 38}, 1837.

\bibitem{TA3} Tsori, Y.; Andelman, D. {\it Macromolecules} {\bf
2003}, {\it 36}, 8560.

\bibitem{tash} Tsori, Y.; Andelman, D. ; Sivaniah, E.; Hashimoto, S. {\it
Macromolecules} {\bf 2005}, {\it 38}, 7193.

\bibitem{GHF} Riise, B. L.; Fredrickson, G. H.; Larson, R. G.; Pearson, D. S.
{\it Macromolecules} {\bf 1995}, {\it 28}, 7653. Koppi, K. A.; Tirrell, M.;
Bates, F. {\it Phys. Rev. Lett.} {\bf 1993}, {\it 70}, 1449.

\bibitem{AH93} Amundson, K.; Helfand, E.; Quan, X; Smith,  S. D. {\it
Macromolecules} {\bf 1993}, {\it 26}, 2698.

\bibitem{russell2} Thurn-Albrecht, T.; Schotter, J.; K\"{a}stle, G.
A.; Emley, N.; Shibauchi, T.; Krusin-Elbaum, L.; Guarini, K.;
Black, C. T.; Tuominen, M. T.; Russell, T. P. {\it Science} {\bf
2000}, {\it 290}, 2126.

\bibitem{TA2} Tsori, Y.; Tournilhac, F.; Andelman, D.; Leibler, L.
{\it Phys. Rev. Lett.} {\bf 2003}, {\it 90}, 145504. Tsori, Y.;
Tournilhac, F.; Leibler, L. {\it Macromolecules} {\bf 2003}, {\it 36},
5873. Tsori, Y.;
Andelman, D. {\it Macromolecules} {\bf 2002}, {\it 35}, 5161.


\bibitem{muthu2} Ashok, B.; Muthukumar, M.; Russell, T. P. {\it J.
Chem. Phys.} {\bf 2001}, {\it 115}, 1559. Pereira, G. G.; Williams,
D. R. M. {\it Macromolecules} {\bf 1999}, {\it 32}, 8115.

\bibitem{russell3} Wang, J.-Y.; Xu, T.; Leiston-Belanger, L. S.; Gupta, S.;
Russell, T. P. {\it Phys. Rev. Lett.} {\bf 2006}, {\it 96}, 128301.

\bibitem{schick} Lin, C.-Y.; Schick, M.; Andelman, D. {\it Macromolecules} {\bf
2005}, {\it 38}, 5766. Tsori, Y.; Andelman, D.; Lin, C.-Y.; Schick, M. {\it
Macromolecules} {\bf 2006}, {\it 39}, 289.

\bibitem{krausch1} B\"{o}ker, A.; Knoll, A.; Elbs, H.; Abetz, V.;
M\"{u}ller, A. H. E.; Krausch, G. {\it Macromolecules} {\bf 2002},
{\it 35}, 1319. B\"{o}ker, A.; Elbs, H.; H\"{a}nsel, H.; Knoll, A.;  
Ludwigs, S.; Zettl, H.; Zvelindovsky, A. V.; Sevink, G. J. A.; Urban, V.; 
Abetz, V.; M\"{u}ller, A. H. E.; Krausch, G.
{\it Macromolecules} {\bf 2003}, {\it 36}, 8078.

\bibitem{krausch2} B\"{o}ker, A.; Schmidt, K.; Knoll, A.;  
Zettl, H.; H\"{a}nsel, A.; Urban, V.; Abetz, V.; Krausch, G.
{\it Polymer} {\bf 2006}, {\it 47}, 849.

\bibitem{russell4}  Xu, T.; Zvelindovsky, A. V.; Sevink, G. J. A.; Gang, O.;
Ocko, B.; Zhu, Y. Q.; Gido, S. P.; Russell, T. P.
{\it Macromolecules} {\bf 2004}, {\it 37}, 6980.

\bibitem{russell5}  Xu, T.; Goldbach, J. T.; Russell, T. P. {\it
Macromolecules} {\bf 2003}, {\it 36}, 7296.

\bibitem{matsen} Matsen, M. W. {\it Macromolecules} {\bf 2006}, {\it 39}, 5512.

\bibitem{zvelin1}  Xu, T.; Zvelindovsky, A. V.; Sevink, G. J. A.; Lyakhova, K.
S.; Jinnai, H.; Russell, T. P. {\it Macromolecules} {\bf 2005}, {\it 38}, 10788.

\bibitem{zvelin2} Zvelindovsky, A. V.; Sevink, G. J. A.
{\it J. Chem. Phys.} {\bf 2005}, {\it 123}, 074903.

\bibitem{russell7} Morkved, T. L.; Lu, M.; Urbas, A. M.; Ehrichs, E. E.; Jaeger, H. M.;
Mansky, P.; Russell, T. P. {\it Science} {\bf 1996}, {\it 273}, 931.

\bibitem{TTL} Tsori , Y.; Tournilhac, F.; Leibler, L. {\it Nature}
{\bf 2004}, {\it 430}, 544.



\bibitem{russell6} Xu, T.; Hawker, C. J.; Russell, T. P. {\it
Macromolecules} {\bf 2005}, {\it 38}, 2802.

\bibitem{matsen_prl} Matsen, M. W. {\it Phys. Rev. Lett.} {\bf 2006}, {\it 95}, 258302.

\bibitem{pohl} Pohl, H. A. {\it Dielectrophoresis}; (Cambridge University Press:
Cambridge, UK, 1978).

\end{thebibliography}
\end{document}